# Diffraction as scattering under the Born approximation

*Neha Goswami and Gabriel Popescu*

*Quantitative Light Imaging Laboratory*

*Beckman Institute of Advanced Science and Technology*

*University of Illinois at Urbana-Champaign, Illinois 61801, USA*

*\*gpopescu@illinois.edu*

**Abstract**

Light diffraction at an aperture is a basic problem that has generated a tremendous amount of interest in optics. Some of the most significant diffraction results are the Fresnel-Kirchhoff and Rayleigh-Sommerfeld formulas. These theories are based on solving the wave equation using Green's theorem and result in slightly different expressions depending on the particular boundary conditions employed. In this paper, we propose another approach for solving the diffraction by a thin screen, which includes apertures, gratings, transparencies etc. We show that, applying the first order Born approximation to 2D objects, we obtain a general diffraction formula, without angular approximations. We discuss several common approximations and place our results in the context of existing theories.

According to Born and Wolf, "Diffraction problems are amongst the most difficult ones encountered in optics" [1]. During a period when Newton's corpuscular theory of light [2] was generally accepted, Huygens' incredible intuition provided new insights into the propagation of optical fields [3]. Later on, as Young [4] and Fresnel [5] cemented the wave theory of light, Huygens' principle was placed in a more rigorous mathematical formalism (for a collections of memoirs by Huygens, Young, and Fresnel, one can consult Ref. [6]). Informative reviews in historical context of the scalar diffraction theory can be found in [1] (Chap. VII) and [7] (Chap. 3). One common problem is the diffraction of a spherical wave at an aperture (see Fig. 1). In this case, the Fresnel-Kirchhoff and Rayleigh-Sommerfeld formulas can be summarized as (see [7], Chap. 3 for derivations)

$$U_1(\mathbf{r}) = \frac{A}{i\lambda} \iint_a \frac{e^{i\beta_0 |(\mathbf{r}'_\perp,0)-\mathbf{r_0}|}}{|(\mathbf{r}'_\perp,0)-\mathbf{r_0}|} \frac{e^{i\beta_0 |\mathbf{r}_\perp - \mathbf{r}'_\perp, z|}}{|\mathbf{r}_\perp - \mathbf{r}'_\perp, z|} f(\theta) d^2\mathbf{r}'_\perp \qquad [1]$$

where the $\perp$ subscript indicates transverse vectors, the position vectors are shown in Fig. 1a, and $f(\theta)$ is the angular dependence of the secondary point sources on the aperture, which can be written as

$$f(\theta) = \begin{cases} \left(\dfrac{\cos\theta_1 - \cos\theta_0}{2}\right) \text{ for Fresnel – Kirchoff formula} \\ \cos\theta_1 \text{ for first Rayleigh – Sommerfeld formula} \\ -\cos\theta_0 \text{ for second Rayleigh – Sommerfeld formula} \end{cases}. \qquad [2]$$

Note that these expressions give the same result for small angles of incidence and diffraction, but can result in drastically different solution for large angles.

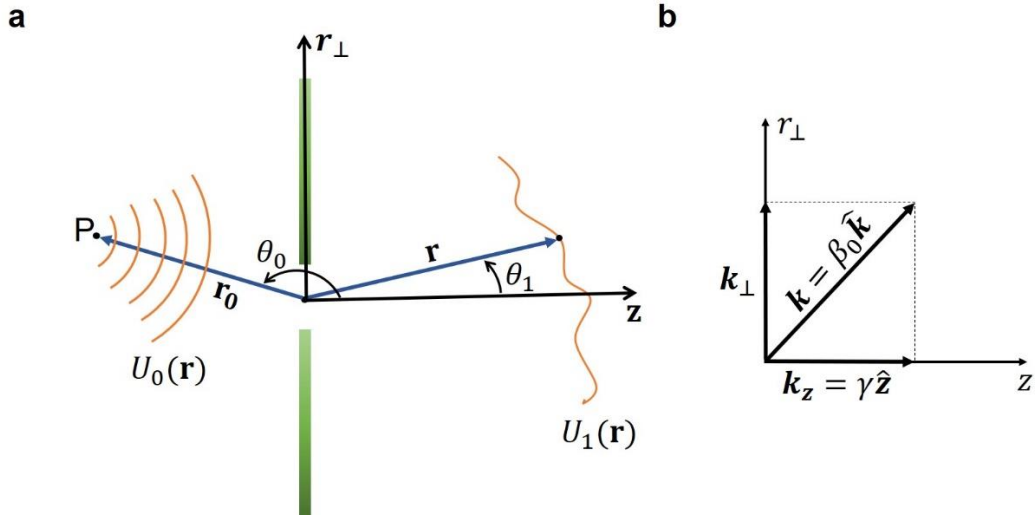

**Figure 1. Diffraction by an aperture:** a) P is the illuminating point source with position vector, $\mathbf{r_0}$, making an angle $\theta_0$ with the outward normal to aperture surface. $U_0(\mathbf{r})$ is the incident field at the aperture and $U_1(\mathbf{r})$ is the diffracted field at observation point with position vector $\mathbf{r}$ making an angle $\theta_1$ with the outward normal to aperture surface. b) Wavevector, with its transverse and longitudinal component notations, as used in text.

Here, we derive the general formula for the diffracted field generated by an arbitrary incident field at a screen, which includes the aperture above. As a particular case, we solve the diffraction problem of the spherical wave at an aperture and we express the field without angular approximations. We start with the inhomogeneous Helmholtz equation,

$$\left[\nabla^2 + \beta^2(\mathbf{r})\right] U(\mathbf{r};\omega) = 0 \qquad [3]$$

where, $\beta(\mathbf{r}) = n(\mathbf{r})\omega/c$ is the wavenumber, with $n$ the inhomogeneous refractive index. Equation 3 can be separated into a homogenous and source term, namely

$$(\nabla^2 + \beta_0^2) U(\mathbf{r};\omega) = -\left[n^2(\mathbf{r}) - 1\right]\beta_0^2 U(\mathbf{r};\omega) \qquad [4]$$

Figure 2a shows the diffraction geometry. We consider the incident field, $U_0$, impinging on a transparent screen, infinitesimally thin, such that the refractive index distribution can be modelled as (Fig 2b)

$$n^2(\mathbf{r}) - 1 = \left[n^2(\mathbf{r}) - 1\right] \Pi\left(\frac{z}{\Delta z}\right) \qquad [5]$$

where $\Pi$ is the rectangular function of width $\Delta z$. For an infinitesimally thin object (screen), $n^2(\mathbf{r}) - 1 \simeq \left[n^2(\mathbf{r}_\perp) - 1\right] \Delta z \delta z$, where $\delta$ is Dirac's function and $\mathbf{r}_\perp = (x, y)$. We define the transmission function of the diffracting screen to be of the form,

$$t(\mathbf{r}_\perp) = \beta_0 \left[n^2(\mathbf{r}_\perp) - 1\right] \Delta z \qquad [6]$$

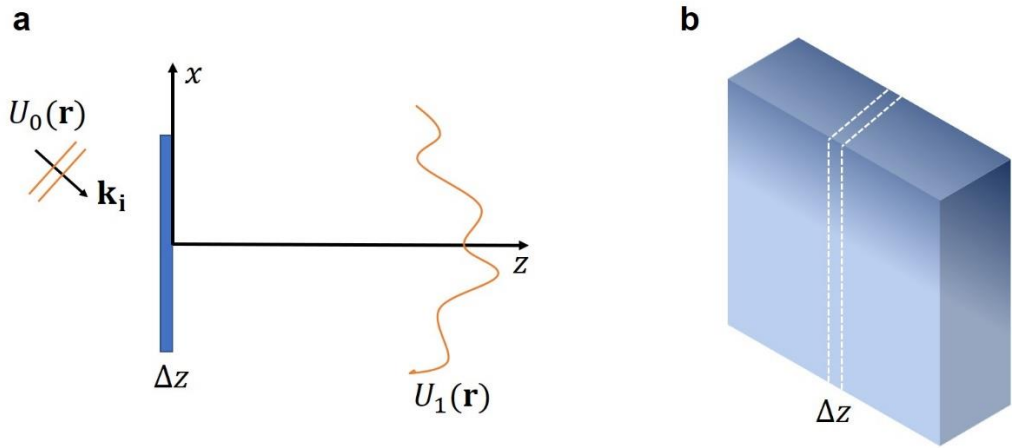

**Figure 2. Diffraction at a thin screen:** a. Plane wave, $U_0(\mathbf{r})$, propagating with a wavevector, $\mathbf{k}_i$, is incident on a thin diffracting object. Each point inside the aperture acts as a secondary point source, emitting spherical waves that sum up to form a diffraction pattern at an observation plane. b. The diffracting screen is modelled as a slice of infinitesimal width $\Delta z$ in the z direction.

The Helmholtz equation becomes,

$$(\nabla^2 + \beta_0^2)U(\mathbf{r};\omega) = -\beta_0 \delta(z) t(\mathbf{r}_\perp) U(\mathbf{r};\omega) \quad [7]$$

The solution of Eq. 7, $U$, is the sum of the incident field, $U_0$, and diffracted field, $U_1$, $U = U_0 + U_1$, with $U_0$ satisfying the homogenous wave equation. We consider that the thin screens involved in common diffraction problems satisfy the first-order Born approximation, which allows us to replace $U$ on the right hand side of Eq. 7 with $U_0$.

Thus, solving equation 7 for $U_1$ in k-domain, we obtain

$$U_1(\mathbf{k};\omega) = \frac{\beta_0}{k^2 - \beta_0^2}\left[t(\mathbf{k}_\perp) \circledv_\mathbf{k} U_0(\mathbf{k};\omega)\right], \quad [8]$$

where $\circledv_\mathbf{k}$ stands for the convolution integral over $\mathbf{k}$ (we use the same notation for a function and its Fourier transform, but carry the arguments explicitly to avoid any confusions). Taking the inverse Fourier Transform with respect to $k_z$, we use the fact that the term $1/(k^2 - \beta_0^2)$ yields an inverse Fourier Transform for the outgoing wave of the form $i\dfrac{e^{i\gamma(\mathbf{k}_\perp)z}}{2\gamma(\mathbf{k}_\perp)}$, where $\gamma^2(\mathbf{k}_\perp) = \beta_0^2 - k_\perp^2$ (see, e.g., [8]). As a result, Eq. 8 can be re-written as

$$U_1(\mathbf{k}_\perp, z;\omega) = i\frac{\beta_0 e^{i\gamma(\mathbf{k}_\perp)z}}{2\gamma(\mathbf{k}_\perp)} \circledv_z \left[t(\mathbf{k}_\perp)\delta(z) \circledv_{\mathbf{k}_\perp} U_0(\mathbf{k}_\perp, z;\omega)\right]$$

$$= i\frac{\beta_0 e^{i\gamma(\mathbf{k}_\perp)z}}{2\gamma(\mathbf{k}_\perp)} \circledv_z \left[\delta(z) t(\mathbf{k}_\perp) \circledv_{\mathbf{k}_\perp} U_0(\mathbf{k}_\perp, 0;\omega)\right], \quad [9]$$

where we used the property of the delta-function that yields

$$\delta(z)U_0(\mathbf{k}_\perp, z; \omega) = U_0(\mathbf{k}_\perp, 0; \omega)\delta(z).$$

Finally, using the convolution property of a delta-function [9], Eq. 9 simplifies to

$$U_1(\mathbf{k}_\perp, z; \omega) = i\beta_0 \left[ t(\mathbf{k}_\perp) \circledS_{\mathbf{k}_\perp} U_0(\mathbf{k}_\perp, 0; \omega) \right] \frac{e^{i\gamma(\mathbf{k}_\perp)z}}{2\gamma(\mathbf{k}_\perp)}. \qquad [10]$$

Equation 10 represents a very general diffraction formula in the angular spectrum representation, i.e., the $(\mathbf{k}_\perp, z)$ domain, which recovers the well-known result (see, e.g., Ref. 7, Chap. 3, and Fig. 1b). This representation exhibits an interesting feature in that the only z-dependence comes from the phase term $e^{i\gamma(\mathbf{k}_\perp)z}$, which makes predicting the diffracted field distribution at various planes $z = z_1, z_2$, etc., particularly easy.

Bringing the result into the spatial domain, by taking the inverse Fourier transform of Eq. 10 with respect to $\mathbf{k}_\perp$, we recover the Huygens-Fresnel formula, namely,

$$U_1(\mathbf{r}_\perp, z; \omega) = \beta_0 A(\omega) \left[ U_0(\mathbf{r}_\perp, 0; \omega) t(\mathbf{r}_\perp) \circledS_{\mathbf{r}_\perp} \frac{e^{i\beta_0 r}}{r} \right]. \qquad [11]$$

Note that assuming the incident field a plane wave along *z*, $U_0$ vanishes from Eq. 11. Various approximation commonly encountered in practice can be obtained easily from Eq. 10 by invoking different degrees of small-angle assumptions (Fig. 3).

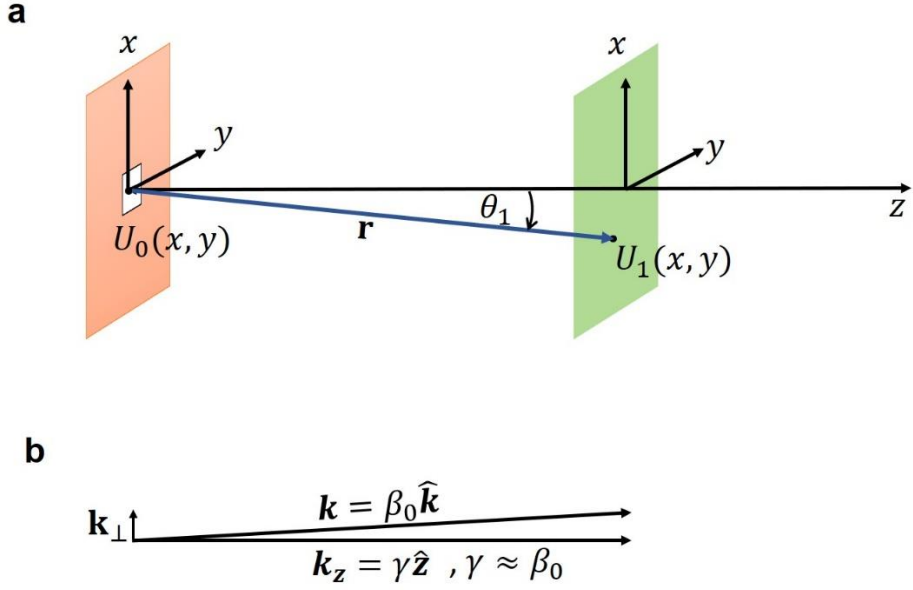

**Figure 3. Approximations:** a. Diffracting aperture is placed at z=0 and observation plane is at z. Field at the aperture and observation is $U_0(x,y)$ and $U_1(x,y)$ respectively, **r** is the position vector of the observation point and $\theta_1$ is the angle of observation. b. For small angle approximations or at sufficiently large distance z, $\gamma \approx \beta_0$.

First, if we approximate $\gamma \simeq \beta_0$ in the amplitude term of Eq. 10, we obtain,

$$U_1(\mathbf{k}_\perp, z; \omega) \simeq \frac{i}{2} A(\omega) t(\mathbf{k}_\perp) e^{i\gamma(\mathbf{k}_\perp)z}. \qquad [12]$$

Equation 12 is the angular spectrum propagation approximation, which simplifies the field propagation significantly. However, Eq. 12 is a fairly accurate representation of the field for most applications, as it keeps the phase term intact. The next, coarser approximation is due to Fresnel, which is obtained by approximating $\gamma$ in the phase term as $\gamma \simeq \beta_0 \left(1 - \frac{k_\perp^2}{2\beta_0^2}\right)$.

Substituting this in Eq. 12

$$U_1(\mathbf{k}_\perp, z; \omega) = \frac{i}{2} A(\omega) t(\mathbf{k}_\perp) e^{i\beta_0 \left(1 - \frac{k_\perp^2}{2\beta_0^2}\right) z}. \qquad [13]$$

In the spatial domain, the Fresnel approximation gives the well-known result

$$U_1(\mathbf{r}_\perp, z; \omega) = \frac{i}{2} A(\omega) e^{i\beta_0 z} \left[ e^{i\beta_0 \left(\frac{r_\perp^2}{2z}\right)} \circledast_{\mathbf{r}_\perp} t(\mathbf{r}_\perp) \right] \qquad [14]$$

Finally, the coarsest approximation is due to Fraunhofer, which, for even smaller angles of diffraction, allows us to neglect quadratic terms in the convolution in Eq. 14. The Fraunhofer approximation yields the diffracted field being expressed as the Fourier transform of the transmission function, namely

$$U_1(x, y, z; \omega) = \frac{i}{2} A(\omega) e^{i\beta_0 z} e^{i\beta_0 \left(\frac{x^2 + y^2}{2z}\right)} t(k_x, k_y)$$
$$k_x = \beta_0 x / z, \quad k_y = \beta_0 y / z \qquad [15]$$

Next, we apply the Born approximation formalism to the classical problem of the spherical wave diffracting at an aperture (Fig. 1a), which was studied by Rayleigh, Fresnel, Kirchhoff and Sommerfeld. In order to solve for the incident field generated by an arbitrary point source at an aperture (Fig. 1a), we consider the source as $\delta(\mathbf{r}) = \delta(\mathbf{r} - \mathbf{r_0})$ driving the wave equation,

$$\nabla^2 U_0(\mathbf{r}) + \beta_0^2 U_0(\mathbf{r}) = \delta(\mathbf{r} - \mathbf{r_0}). \qquad [16]$$

In Eq. 16, the field solution, $U_0$, represents the incident field, which can be used to calculate the diffracted field $U_1$ via Eq. 10. In the $\mathbf{k}$ domain, $U_0$ can be obtained at once,

$$U_0(\mathbf{k}) = \frac{-e^{-i\mathbf{k}\cdot\mathbf{r_0}}}{k^2 - \beta_0^2} \qquad [17a]$$

which in the spatial domain yields a shifted spherical wave,

$$U_0(\mathbf{r}) = \frac{e^{i\beta_0|\mathbf{r}-\mathbf{r_0}|}}{|\mathbf{r}-\mathbf{r_0}|} \qquad [17b]$$

Thus, the diffracted field can be obtained using the general solution in Eq. 11, namely,

$$U_1(\mathbf{r}) = \iint_a \frac{e^{i\beta_0|(\mathbf{r}'_\perp,0)-\mathbf{r_0}|}}{|(\mathbf{r}'_\perp,0)-\mathbf{r_0}|} \frac{e^{i\beta_0|\mathbf{r}_\perp-\mathbf{r}'_\perp,z|}}{|\mathbf{r}_\perp-\mathbf{r}'_\perp,z|} d^2\mathbf{r}'_\perp \qquad [18]$$

Changing the notations (as shown in Fig. 4) to those used in Ref. [7], and using the relationships,

$\mathbf{r_0} = (x_2, y_2, z_2)$, $\mathbf{r}'_\perp = (x_1, y_1)$, $\mathbf{r} = (x_0, y_0, z_0)$, $\mathbf{r}_\perp = (x_0, y_0)$,

$r_{21} = \sqrt{(x_2-x_1)^2 + (y_2-y_1)^2 + z_2^2} = |(\mathbf{r}'_\perp,0)-\mathbf{r_0}|$, $r_{01} = \sqrt{(x_0-x_1)^2 + (y_0-y_1)^2 + z_0^2} = |\mathbf{r}_\perp - \mathbf{r}'_\perp, z|$

Eq. [18] transforms to

$$U(x_0, y_0, z_0) = \frac{A}{i\lambda} \iint_a \frac{e^{i\beta(r_{21}+r_{01})}}{r_{21}r_{01}} ds \qquad [19]$$

which is consistent with Ref. [7], with inclination factor $f(\theta)=1$.

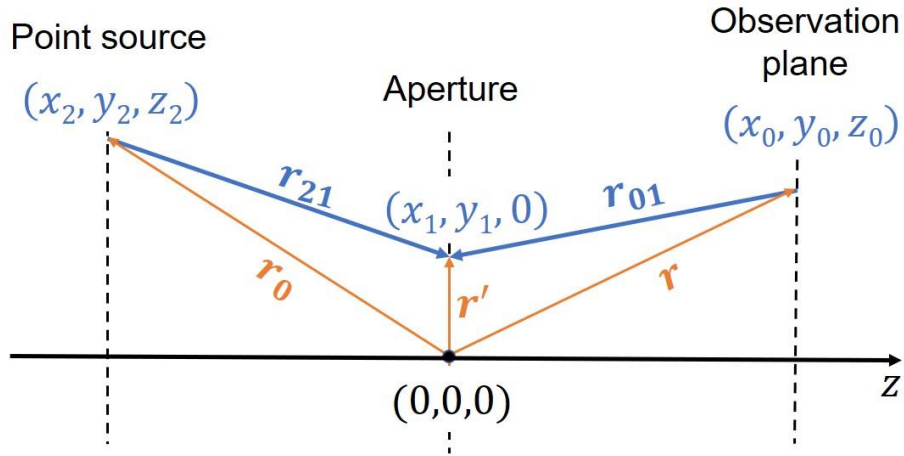

**Figure 4. Notations:** In blue, we show the coordinate notations that convert our result into an identical form with that in Ref. [7].

In summary, we presented an alternative approach to easily derive the expression for diffraction by a thin object. The starting point in this derivation is the realization that scattering and diffraction are fundamentally driven by the same interaction of light with inhomogeneous media. Traditionally, "scattering" refers to interaction with 3D objects, while "diffraction" generally describes the light emerging from 2D objects such as apertures, thin gratings and screens. However, applying the scattering Born approximation to thin 3D objects, we showed that the general formulation for diffraction can be obtained without angular approximations. Our result is consistent with the three classical formulas described in Eqs. 1-2.

Traditionally, the field of light imaging, diffraction, and Fourier Optics, have been mainly concerned with the 2D problem and captured mostly an engineering audience. On the other hand, light scattering, dealing with the 3D interaction, seems to be mostly driven by physicists. We hope that revisiting this classical problem with a unifying formalism of scattering and diffraction may help bring closer the fields of optical imaging and scattering, which are in essence different descriptions on the same light-matter interaction phenomenon.

**Funding:** Authors would like to acknowledge the following grants: National Institutes of Health (R01GM129709, R01CA238191) and National Science Foundation (0939511, 1450962, 1353368) awarded to G. P.

**Disclosures:** The authors declare no competing interests.